\documentclass{article}
\usepackage{graphicx} 
\usepackage{amsfonts} 
\usepackage{mathptmx}       
\usepackage{helvet}         
\usepackage{courier}        
\usepackage{amsmath,bm}
\usepackage{amssymb}
\usepackage{subfigure}
\usepackage{blindtext}
\usepackage{apacite}
\usepackage{natbib}
\usepackage{type1cm}        
\usepackage{makeidx}         
\usepackage{graphicx}        
\usepackage{multicol}        
\usepackage[bottom]{footmisc}
\usepackage{colortbl,booktabs}
\usepackage{tablefootnote}
\usepackage{hyperref}

\title{Unfolding the Network of Peer Grades: A Latent Variable Approach}
\author{Giuseppe Mignemi, Yunxiao Chen and Irini Moustaki}
\date{}

\begin{document}

\maketitle

\begin{abstract}
Peer grading is an educational system in which students assess each other's work. It is commonly applied under Massive Open Online Course (MOOC) and offline classroom settings. With this system, instructors receive a reduced grading workload, 
and students enhance their understanding of course materials by grading others' work. 
Peer grading data have a complex dependence structure, for which all the peer grades may be dependent. This complex dependence structure is due to a network structure of peer grading, where each student can be viewed as a vertex of the network, and each peer grade serves as an edge connecting one student as a grader to another student as an examinee. This paper introduces a latent variable model framework for analyzing peer grading data and develops a fully Bayesian procedure for its statistical inference. This framework has several advantages. 
First, when aggregating multiple peer grades, the average score and other simple summary statistics fail to account for grader effects and, thus, can be biased. The proposed approach produces more accurate model parameter estimates and, therefore, more accurate aggregated grades, by modeling the heterogeneous grading behavior with latent variables. Second, the proposed method provides a way to assess each student's performance as a grader, which may be used to identify a pool of reliable graders or generate feedback to help students improve their grading. Third, our model may further provide insights into the peer grading system by answering questions such as whether a student who performs better in coursework also tends to be a more reliable grader. Finally, thanks to the Bayesian approach,  uncertainty quantification is straightforward when inferring the student-specific latent variables as well as the structural parameters of the model. 
The proposed method is applied to two real-world datasets.

\end{abstract}


\section{Introduction}
\label{Intro}
\begin{figure}
    \centering
    \includegraphics[scale=0.25]{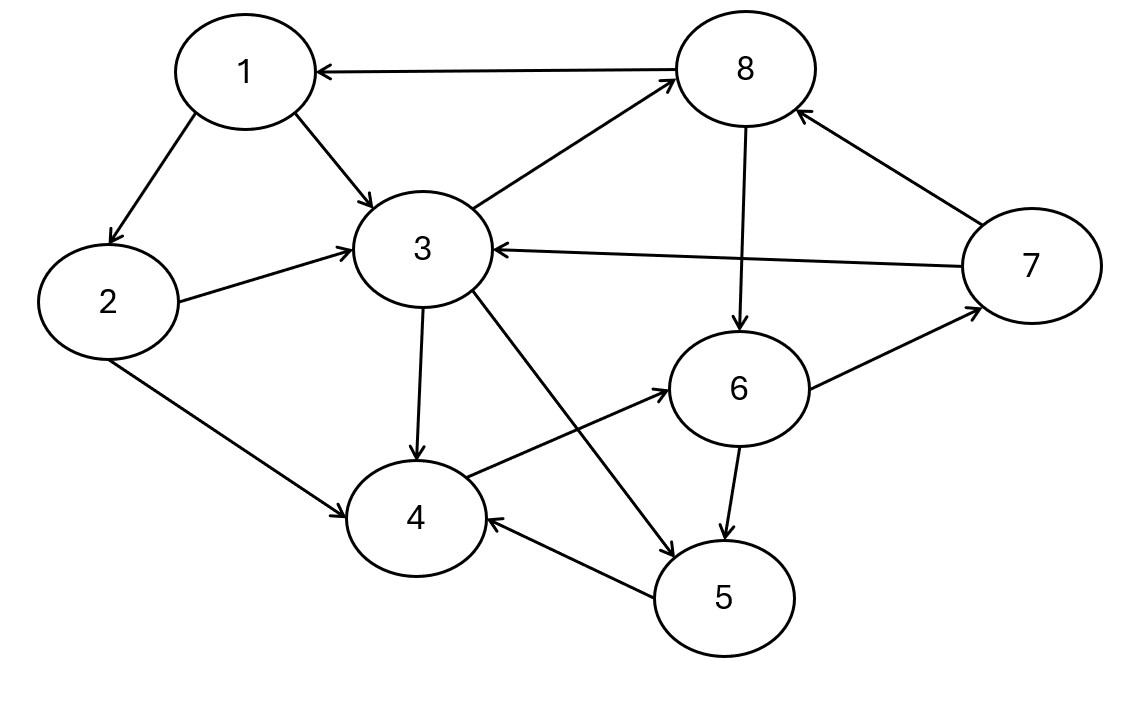}
    \caption{ Network diagram representing the network structure of peer grading data. Each circle is a vertex of the network and represents a student. The arrows are the peer grades, which serve as edges connecting two students; their direction indicates whether the student receives or gives the grade.}
    \label{fig_network_1}
\end{figure}

Peer grading, also known as peer assessment, is a system of formative assessment in education whereby students assess and give feedback on one another's work. It substantially reduces teachers' burden for grading and improves students' understanding of the subject and critical thinking \citep{Yin_2022, Panadero_2019}. Consequently, it is widely used in many educational settings, including massive open online courses (MOOCs; \citealp{Gamage_2021}), large university courses \citep{Double_2016}, and small classroom settings \citep{Sanchez_2017}. In a peer grading system, each student's work is assigned, often randomly, among several other students who act as graders or raters. Due to the design of this system, peer grading data have a different structure from traditional rating data, which also consists of students' grades from graders.
For traditional rating data, the students, whose work is evaluated, cannot serve as graders, which leads to a relatively simple data structure. On the other hand, peer grading data has a network structure where all the peer grades may be dependent. Each student can be viewed as a network vertex, and each peer grade serves as an edge connecting two students -- a grader and an examinee; see Figure \ref{fig_network_1} for a visual illustration of such a network structure.

A simple peer grading system aggregates the peer grades using a straightforward method like the mean or median to derive a final grade for each student's work \citep{Sajjadi2015, Reily2009}. This conventional method does not consider the heterogeneity among the graders. Some graders may exhibit systematic biases and tend to assign higher or lower grades than their peers when assessing the same work. Graders may also exhibit varying levels of reliability; while some maintain consistent grading standards, others may give erratic grades that lack a consistent standard. Furthermore, when the data involve multiple formative assessments for each student, a more accurate grade may be derived by borrowing information across assessments. Finally, monitoring how students perform as graders is often helpful, as it provides an opportunity to reward the best-performing graders and offer feedback to help those who need improvement. Different methods have been developed to mitigate grader bias and improve peer assessment reliability; see  \cite{Alqassab2023} for a review. Depending on whether instructors' scores are needed in method training, they can be classified as supervised and unsupervised learning methods. Supervised learning methods utilize instructors' scores to train a function that maps multiple peer grades to an aggregated grade that mimics the instructor's score \citep{Namanloo2022, Xiao2020}. On the other hand, unsupervised learning methods try to find an aggregation rule only based on peer grades without access to instructors' scores. Unsupervised learning is typically performed by employing latent-variable-based measurement models \citep[e.g.,][]{Han2018,Piech2013,Xu2021}, which are closely related to models for traditional rating data. 
As will be explained in the sequel, they make an independence assumption that is also adopted in the latent variable models for traditional rating data. However, as peer grading data have a complex network structure, this independence assumption is likely over-simplified, leading to suboptimal performance. 

Many latent variable models have been proposed for traditional rating data, including the facet model \citep{linacre1989many} and its extensions \citep{Uto_2020,Uto_2021}, the hierarchical rater models \citep{casabianca2016hierarchical, decarlo2011, Molenaar_2021,nieto2019,Patz2002}, the rater bundle model
\citep{wilson2001rater}, and the generalized rater model \citep{wang2014item}. These models introduce rater-specific parameters to model the rater effects in the data. When having many raters, these rater-specific parameters are treated as random effects (i.e., latent variables) and further assumed to be independent of the examinee-specific latent variables that are used to model examinee performance. These assumptions are also made in the existing latent variable models for peer grading data \citep[][]{Han2018,Piech2013,Xu2021}. However, we note that the assumption about the independence between the rater-specific latent variables and examinee-specific latent variables does not hold for peer grading data, as the same students are both examinees and raters, and the characteristics of the same student as a rater and those as an examinee are naturally correlated. Ignoring such dependence can result in model misspecification and substantial information loss. To our knowledge,
no rater model in the literature accounts for such a dependence structure.

We fill this gap by proposing an unsupervised latent variable model for peer grading data. The proposed model jointly analyzes peer grades for multiple assessments and produces more accurate aggregated grades. It models the student effects with correlated latent variables that capture a student's characteristics as an examinee and a grader, respectively. Unlike the existing latent variable models for peer grading data, the proposed model captures the dependence in data brought by the network structure of peer grades and the dual roles of each student as an examinee and a rater.

Due to the complex dependence structure under the proposed model, its marginal likelihood involves a very high-dimensional integral with respect to all the student-specific latent variables that can hardly be simplified.   Thus, solving the maximum likelihood estimator is computationally infeasible, and consequently, frequentist inference based on the marginal likelihood is a challenge. We develop a fully Bayesian approach for drawing statistical inferences to overcome the computational challenge. With this approach, uncertainty quantification is straightforward when inferring the student-specific latent variables as well as the structural parameters of the model. However, its computation is still non-trivial due to the presence of a large number of latent variables and a complex network structure. To solve this, we use a No-U-Turn Hamiltonian Monte Carlo sampler \citep{Hoffman_2014}, which produces efficient approximate samples from the posterior distribution. 

Besides the traditional rater models, the proposed framework is closely related to cross-classified random effects models \citep{raudenbush1993crossed,goldstein1994multilevel,rasbash1994efficient}, an extension of standard multilevel models for non-hierarchical data that have cross-classified structures. These models have received wide applications for evaluating measurement reliability, including in generalizability theory \citep{Brennan2001, Brennan2010}. Our data involve three crossed factors - the examinees, the graders, and the assessments, and the proposed model decomposes each peer grade based on these three factors. However, our model allows the latent variables (i.e., random effects) associated with the crossed factors (examinees and raters) to be correlated to account for the special design of peer grading. In contrast, a standard cross-classified random effects model assumes the random effects associated with different crossed factors to be independent. Introducing such dependence among the latent variables substantially increases the complexity of the model and its inferences. Our model also has close connections with several latent variable models concerning dyadic data, including social relations models \citep[e.g.,][]{kenny1984social,nestler2016restricted, Nestler2017, Nestler2020,warner1979new} and the dyadic item response theory (IRT) model \citep{Gin_2020}, where the dyadic IRT model extends the social 
relations models by incorporating an IRT measurement model. Peer grading data can be viewed as a special type of dyadic data, where each dyad involves an examinee and a grader, and the dyads are formed by random assignment. However, our model differs substantially from the existing social relations models in how latent variables are modeled and interpreted. The traditional social relations models focus on inferring the causes and consequences of interpersonal perceptions and judgments. In contrast, the current analysis focuses on measuring some latent traits that are concerned in the application of peer grading (e.g., examinee performance and rater reliability). As a result, the existing social relations models are unsuitable for the current application. 

The rest of the paper is organized as follows. Section~\ref{sec:model} proposes a latent variable model framework for peer grading data, within which specific models are discussed. Two real data applications are given in Section~\ref{sec:real}. Section~\ref{sec:discussion} discusses advantages, limitations, and future directions. The appendix includes extensions of the proposed model, technical details, and additional simulated examples.

\section{Proposed Model}\label{sec:model}

\subsection{Problem Setup}
Consider $N$ students who receive $T$ assessments. Each student $i$'s work on assessment $t$ is randomly assigned to a small subset of other students to grade their work. We denote this subset as $S_{it}$, which is a subset of $\{1, ..., i-1, i+1, ..., N\}$. Each grader $g \in S_{it}$ gives this work a grade $Y_{igt}$, following certain scoring rubrics. For simplicity, we consider the case when $Y_{igt}$ is continuous. 
It is common, but not required, for the number of graders $|S_{it}|$ to be the same for all students and assessments. An aggregated score is then computed as a measure of student $i$'s performance on the $t$th assessment, often by taking the mean or the median of the peer grades $Y_{igt}, g \in S_{it}$. We note that a simple aggregation rule, such as the mean and the median of the peer grades, fails to account for the grader effect and, thus, may not be accurate enough. 

\subsection{Proposed Model}\label{Main}

\paragraph{Modelling Peer Grade $Y_{igt}$.}
We assume the following decomposition for the peer grade $Y_{igt}$:
\begin{equation}\label{eq:1}
    Y_{igt} = \theta_{it} + \tau_{igt} - \delta_t, \quad i = 1, ..., N, t= 1,\dots, T, g\in S_{it}.
\end{equation}
Here, $\delta_t$  captures the difficulty level of assessment $t$. A larger value of $\delta_t$ corresponds to a more difficult assessment. 
In addition, $\theta_{it}$ represents student $i$'s true score for assessment $t$, and $\tau_{igt}$ is an error attributed to the grader. We assume $\theta_{it}$,  $\tau_{igt}$ and $\delta_t$ to be independent.

\paragraph{Modeling True Score $\theta_{it}$.} 
For each student $i$, we assume that their true scores for different assessments $\theta_{it}, \; t = 1,\dots, T$, are independent and identically distributed (i.i.d.), following a normal distribution 
\begin{equation}\label{eq:truescore}
\theta_{it} \sim N(\alpha_{i}, \eta_{i}^2), 
\end{equation}
where the mean and variance are student-specific latent variables. The latent variable $\alpha_{i}$ captures the student's average performance over the assessments, and the latent variable $\eta_{i}^2$ measures their performance consistency (i.e., the extent to which student's proficiency varies across assessments). This model assumes the true scores fluctuate randomly around the average score $\alpha_{i}$ without a trend. This assumption can be relaxed if we are interested in assessing students' growth over time; see Appendix \ref{Extensions} for a relaxation of this assumption.

\paragraph{Modelling Grader Effect $\tau_{igt}$.} 
Each student $g$ grades multiple assessments from multiple students. We let $H_g = \{(i,t): g \in S_{it}, t = 1, ..., T\}$
be all the work student $g$ grades.
For each student $g$, we assume that $\tau_{igt}$, for all $(i,t) \in H_g$, are i.i.d., following a normal distribution $N(\beta_{g}, \phi_{g}^2)$,
where the mean and variance are student-specific latent variables. The latent variable $\beta_{g}$ may be interpreted as the bias of student $g$ as a grader. For two students $g$ and $g'$ satisfying $\beta_g > \beta_{g'}$, student $g$ will give a higher grade on average than student $g'$ when grading the same work. We say grader $g$ is unbiased when $\beta_g = 0$. Moreover, the latent variable $\phi_{g}^2$ measures the grader's reliability.  A smaller value of $\phi_{g}^2$ implies that the grader provides consistent grades to assessments of similar quality, while a larger value suggests the opposite. 
In other words, when grading
multiple pieces of work with the same true score and assessment difficulty (so that ideally they should receive the same grade), a grader with a small $\phi_g^2$ tends to give similar grades, and thus, the grades are more reliable. In contrast, a grader with a large $\phi_g^2$ tends to give noisy grades that lack consistency. 
We remark that the grader effects $\tau_{igt}$, $t = 1, ..., T$, are assumed to be i.i.d. in the current setting, which means the grading quality remains the same over time. 
 
\paragraph{Joint Modelling of Student-Specific Latent Variables.} The model specification above introduces four latent variables, namely $\alpha_{i}$, $\beta_{i}$, $\eta_{i}^2$, and $\phi_{i}^2$, for each student $i$. These variables allow us to account for the relationship between a student's performance data and grading data as an examinee and a grader. By allowing for dependence between these variables, we can share information and make more informed evaluations of their performance.
We assume that 
$(\alpha_{i}, \beta_{i}, \eta_{i}^2,  \phi_{i}^2)$, where $i = 1, ..., N$ are i.i.d.; we also assume that $(\alpha_{i}, \beta_{i},  \log(\eta_{i}^2), \log(\phi_{i}^2))$ follows a multivariate normal distribution  $N\left(\bm{\mu}, \bm{\Sigma}  \right)$, where $\bm{\mu} = (\mu_1, ..., \mu_4)^\top$ and $\bm{\Sigma} = (\sigma_{mn})_{4\times 4}$. To ensure parameter identifiability, we set  $\mu_1 =  \mu_2 =0$ so that the average score of each assessment (averaged across students and graders) is completely captured by the difficulty parameter $\delta_t$.  There are no constraints on $\mu_3$ and $\mu_4$.

\paragraph{Remarks.}

\begin{figure}
    \centering
    \includegraphics[scale=0.5]{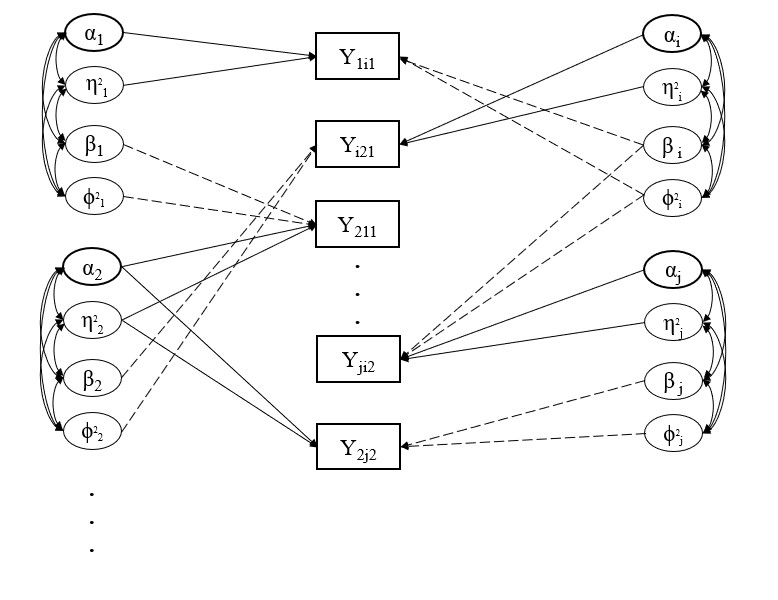}
    \caption{ Path diagram representing the network structure of peer grading data. The latent variables of four independent students are represented as an example. Students' grades, reported in the squared box, refer to two assessments, as the subscripts indicated. The curve double-arrows stand for correlation; the straight (solid and dotted) lines represent the effect of the respective latent variable. For the sake of readability, we prefer to adopt the solid lines for the effect of variables referring to the role of the examinee (i.e., $\alpha, \eta^2$), whereas the dotted lines refer to the effect of the latent variables associated to the role of grader (i.e., $\beta, \phi^2$).}
    \label{fig:path}
\end{figure}

Figure~\ref{fig:path} shows an illustrative path diagram for the proposed model under a simplified setting with $N=4$ students and $T=2$ assessments. Compared with many traditional latent variable models, the current path diagram shows a network structure where the latent variables of different individuals interact with each other. This phenomenon is due to the network structure of peer grading data, where each grade involves two students -- one as the examinee and the other as the grader. 

The proposed model is useful in different ways. First, the model provides a measurement model for the true score of each student $i$'s assessment $t$. By inferring each latent variable, $\theta_{it}$, whose technical details will be discussed in Section~\ref{Bayes}, the grader and assessment effects will be adjusted. Thus, a more accurate aggregated score may be obtained. Second, it allows us to further assess each student's overall performance and consistency as an examinee by inferring $\alpha_{i}$ and $\eta_{i}^2$. Third, the model also provides a measurement model for the characteristics of each student as a grader. Specifically, the bias and reliability of each grader can be assessed by inferring $\beta_{i}$ and $\phi_{i}^2$. Such results can be used to reward the best-performing graders and offer feedback to help those who need improvement. Finally,  the statistical inference of the structural parameters in $\boldsymbol{\Sigma}$ allows us to address substantive questions, such as whether a student who performs better in the coursework tends to be a more reliable grader.  

\subsection{Bayesian Inference}\label{Bayes}

We adopt a fully Bayesian procedure for drawing statistical inference under the proposed model.   

\paragraph{Prior specification.} \label{Prior}
We first specify the prior for the assessment difficulty parameters $\delta_1,\dots,\delta_T$. 
When $T$ is large, we can get reliable estimates of the assessments' population parameters (e.g., the mean and the variance, \citealt{Cao_2008, DeBoeck_2008, Gelman_2006_BA,fox2001}). In such cases, we can use a hierarchical prior specification and assume that 
 $\delta_1,\dots,\delta_T$ are i.i.d. following a specific prior distribution (e.g., a normal distribution) with some hyper-parameters. Then, we set a hyper-prior distribution for the hyper-parameters. When $T$ is small, it is not reasonable to assume to observe a representative sample of assessments, and the estimates at the population level might be very unreliable \citep {DeBoeck_2008}. Therefore, we let each $\delta_t$ have a weakly informative prior distribution of $N(0, 25)$. However, tailored considerations must be made depending on the specific data set, and different prior specifications might be specified \citep{Gelman2013}. 
 
We specify a prior for the parameters $\bm \mu$ and $\bm \Sigma$ in the joint distribution for the student-specific latent variables. Recall that $\mu_1$ and $\mu_2$ are constrained to zero, so no prior is required. As for $\mu_3$ and $\mu_4$, they are assumed to be independent, and each follows a weakly informative normal prior $N(0, 25)$. Finally, 
for the covariance matrix $\bm \Sigma$, we reparameterize it as
\begin{eqnarray}\label{eq:3}
    \bm{\Sigma} &=& \textbf{S}  \bm{\Omega}  \textbf{S}, \nonumber
\end{eqnarray}
where $\textbf{S}=\mbox{diag}(\sqrt{\sigma_{11}},\dots,\sqrt{\sigma_{44}})$ is a $4 \times 4$ diagonal matrix with diagonal entries the standard deviations of $(\alpha_{i}, \beta_{i},  \log(\eta_{i}^2), \log(\phi_{i}^2))$, and $\bm{\Omega} = (\omega_{ij})_{4\times 4} = \textbf{S}^{-1}\bm{\Sigma}\textbf{S}^{-1}$ is the correlation matrix of $(\alpha_{i}, \beta_{i},  \log(\eta_{i}^2), \log(\phi_{i}^2))$.  The prior distribution on $\bm{\Sigma}$ is imposed through the priors on $\textbf{S}$ and $\bm{\Omega}$. For 
$\textbf{S}$, we assume 
$\sqrt{\sigma_{11}},\dots,\sqrt{\sigma_{44}}$ to be i.i.d., each following a half-Cauchy distribution with location $0$ and scale $5$. For the correlation matrix $\bm{\Omega}$, we assume a 
Lewandowski-Kurowicka-Joe (LKJ) prior distribution with shape parameter 1 \citep{LKJ}, which corresponds to the uniform distribution over the space of all correlation matrices.

\paragraph{Model Comparison.}\label{ModelComp}
Several reduced models can be derived under the proposed framework as special cases. For instance, a reduced model may be obtained by constraining $\eta_1^2 = \cdots = \eta_N^2 = \eta^2$, i.e., students' performance consistency as examinee is homogeneous. Another reduced model may be derived by constraining $\phi_1^2 =  \cdots = \phi_N^2 = \phi^2$. An even more simplified model can be obtained by imposing both sets of constraints. Given a dataset, Bayesian model comparison methods may be used to find the best-performing model among the full and the reduced models and, thus, provide insights into the peer grading system and yield more accurate aggregated grades. 

We consider a Bayesian leave-one-out (LOO) cross-validation procedure for model comparison, which concerns the model's prediction performance. For a given dataset and a given model, this procedure computes the expected log point-wise predictive density (elpd; \citealp{Vehtari2017}) to measure the overall accuracy in predicting each data point (i.e., peer grade) based on the rest of the data. More precisely, we define the Bayesian LOO estimate of out-of-sample predictive fit as
\begin{eqnarray}\label{eq:5}
    \mbox{elpd}_{loo} &=& \sum_{t=1}^{T} \sum_{i=1}^{N} \sum_{g \in S_{it}} \log p(Y_{igt}|\textbf{Y}_{-igt}), \nonumber 
\end{eqnarray}
where $\textbf{Y}_{-igt}$ denotes all the observed peer grades except for $Y_{igt}$, and  $p(Y_{igt}|\textbf{Y}_{-igt})$ denotes the conditional probability mass function of $Y_{igt}$ given $\textbf{Y}_{-igt})$ under the fitted Bayesian model.  A model with a higher value of $\mbox{elpd}_{loo}$ is regarded to have better prediction power and, thus, is preferred. In Section \ref{sec:real}, we also report the Watanabe–Akaike information criterion (WAIC), which corrects the expected log point-wise predictive density by adding a penalty term for the effective number of parameters \citep{Vehtari2017}.

\paragraph{Computation.}

As illustrated in Figure~\ref{fig:path}, the proposed model involves a latent space with dimension $4N$ and a complex dependence structure between the observed data and the latent variables. This complex model structure makes its statistical inference computationally a challenge. We use a Markov Chain Monte Carlo (MCMC) algorithm for statistical inference. More specifically,  we adopt the No-U-Turn Hamiltonian Monte Carlo (HMC) sampler \citep{Hoffman_2014}, a
computationally efficient MCMC sampler, and implement it under the Stan programming language. 
Compared with classical MCMC samplers, such as the Gibbs and Metropolis-Hastings samplers, the No-U-Turn HMC sampler uses geometric properties of the target distribution to propose posterior samples. It thus converges faster to high-dimensional target distributions \citep{Hoffman_2014}. 
Further computational details are given in the appendix. 

Regarding the implementation, we use the CmdStan interface \citep{stan} for posterior sampling, which is a command-line interface to Stan that is considerably more efficient than using R as the interface. For all the models, 4 HMC chains are run in parallel for 2,000 iterations, of which the first 1,000 iterations were specified as the burn-in period. We use the rstan R package to analyze the resulting posterior samples, more specifically, it enables us to merge the MCMCs, compute the summary statistics of the posteriors and check the MCMC mixing and convergence. Moreover, the R package loo \citep{Vehtari2017} and Bayesplot \citep{Bayesplot} are used separately for model comparisons and to plot the results, respectively. The computation code used in our analysis, the computational time, and other details on model diagnostics are publically available online\footnote{Our code is available through the link: \url{https://osf.io/v3ucw/?view_only=aad3bc91cbda43cc9e6c490409323839}}.

\subsection{A Related Model}\label{MP}
One of the most well-known approaches to latent variable modeling of peer grading data was proposed by \citealt{Piech2013}. They present three models of increasing complexity, in which the observed score is assumed to be a function of two independent variables: the student's ability (also known as the true score) and the effect of the grader (often considered the error part). This type of decomposition is very common in rater effects models \citep{Martinkova2023,Gwet_Li} and is also assumed in our framework. For comparison purposes, we briefly discuss their more complex model,  which is also considered in Section~\ref{sec:real} and compared with the one we present in Section \ref{Main}. The notation we adopt in presenting their model is consistent with our framework.
They assume that the observed score $Y_{ig}$ is normally distributed with the mean parameter given by the sum of the true score $\theta_i$ and the grader bias $\beta_g$, and the precision parameter being a linear function of the true score of student $g$:
\begin{eqnarray}
    Y_{ig} &\sim& N\left( \theta_i + \beta_g, \frac{1}{\gamma_0+\gamma_1 \theta_g}\right). \nonumber
\end{eqnarray}
The model assumes that the true scores of students,  denoted by $\theta_i$, are independently and identically normally distributed, $\theta_i \sim N(\mu_0, 1/\gamma_2)$, $i=1,\dots,N$. 
In addition, the model assumes that graders' biases denoted by $\beta_g$,  are i.i.d. normally distributed, $\beta_g \sim N(0,1/\gamma_3)$, $g=1,\dots,N$. 

While this model relates to the proposed method, the two have several differences. For example, the model proposed by the authors does not account for the difficulty level of the assignment. Even if they propose to use normalized grades ($z$-scores) to remove any assignment effects, it may still be useful to estimate the difficulty level of the assignment.
Furthermore, the model assumes that the parameters $\theta_i$ and $\beta_i$, which are the same student indexes, are independent. It also imposes a strict constraint on the precision parameter of the observed score. Specifically, it does not allow the precision to vary given the same value of $\theta_g$, and it assumes that the precision is independent of the grader bias $\beta_g$. Finally, the model does not account for the temporal dependency in the presence of multiple assessments. This model, denoted in Section~\ref{sec:real} as PM (i.e., Piech's Model), is compared against the proposed one using real data from multiple and single assessment contexts.

\subsection{Reduced Model for a Single Assessment}\label{CrossSectional}

Some peer grading data only involve a single assessment, as the case for one of our real data examples in Section~\ref{sec:real}. The proposed model can still be applied in that situation, but certain constraints must be imposed for model identification. Details of the Bayesian inference for this model are given in the appendix.

\paragraph{Modelling Peer Grade $Y_{ig}$.}

With only one assessment, the notation for peer grade simplifies to $Y_{ig} = Y_{ig1}$, and its decomposition simplifies to 

\begin{equation}\label{eq:red}
    Y_{ig} = \theta_{i} + \tau_{ig} - \delta, \quad i = 1, ..., N, g\in S_{i},
\end{equation}
where the subscript $t$ is removed from all the notations in \eqref{eq:1}, and the interpretation of the variables remains the same. Due to the lack of multiple assessments, the examinee parameters $\alpha_{i}$ and $\eta_{i}^2$ in the main model, Equation  (\ref{eq:truescore}), can no longer be identified and, thus, are not introduced here.

\paragraph{Modelling Grader Effect $\tau_{ig}$.} 
Each student $g$ grades the assessment of multiple peers. Let  
$H_g = \{i: g \in S_i\}$ be the peers whose work is graded by student $g$. It is assumed that  $\tau_{ig}$, $i \in H_g$, are i.i.d., following a normal distribution $N(\beta_{g}, \phi_{g}^2)$. 
The interpretation of these parameters is the same as in the primary model; see Section \ref{Main}.  

\begin{figure}
    \centering
    \includegraphics[scale=0.5]{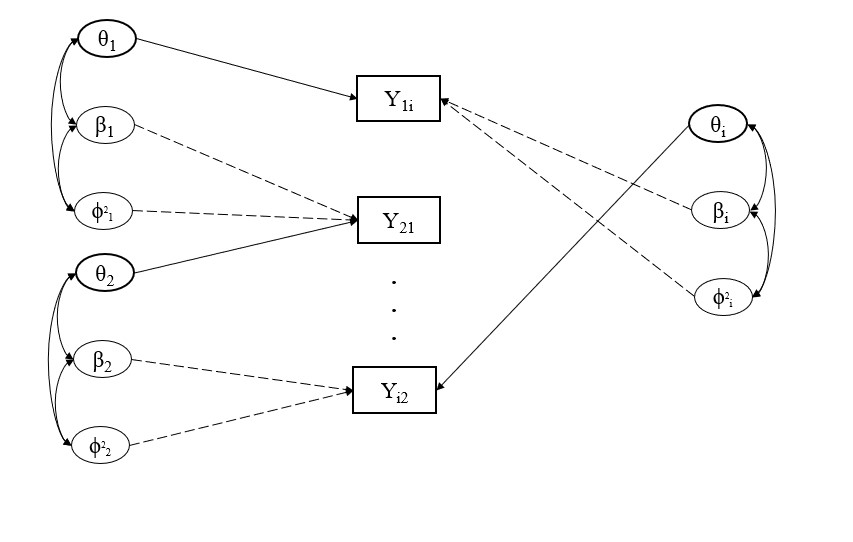}
    \caption{ Path diagram representing the network structure of peer grading data for a single assessment. The latent variables of three independent students are represented as an example. The box indicates the students' grades for a single assessment. The double arrows represent correlation, while the straight (solid and dotted) lines represent the effect of the respective latent variable. The meaning of the arrows is consistent with those of Figure \ref{fig:path}. The solid line represents the effect of the latent variable related to the role of the examinee (i.e., $\theta$). The dotted lines refer to the effect of the latent variables associated with the grader role (i.e., $\beta, \phi^2$).} 
    \label{fig:single}
\end{figure}

\paragraph{Joint Modelling of Student Specific Latent Variables.} 
The reduced model involves three student--specific latent variables  $(\theta_i,\beta_{i},\phi_{i}^2)$. Similar to the main model, 
we assume that $(\theta_i,\beta_{i},\log(\phi_{i}^2))$, $i=1, ..., N$, are i.i.d., each following a multivariate normal distribution
$N\left(\bm{\mu}, \bm{\Sigma}  \right)$, where $\bm{\mu} = (\mu_1, \mu_2, \mu_3)^\top$ and $\bm{\Sigma} = (\sigma_{ij})_{3\times 3}$. Similar to the main model, we constrain 
$\mu_1 = \mu_2 = 0$, while keep $\mu_3$ freely estimated. Figure~\ref{fig:single} gives an illustrative path diagram for this reduced model with $N=3$ students.

\section{Real Data Examples}\label{sec:real}

Two real-world applications referring to single- and multiple-assessments settings are considered here. Various models are compared for each data set, and the one that exhibits the best predictive performance is used for inference.

\subsection{Multiple-Assessments Setting}\label{subsec:real_Long}
The peer grading data is from \cite{Zong2021}. In this data, 274 American undergraduate students taking a Biology course completed four double-blind peer gradings throughout the course ($N=274, T=4$). 
The assessments had a similar format, and the online peer reviewing system managed the submission and peer grading procedures \textit{SWoRD/Peerceptiv} \citep{Patchan2016}. Students' mean age was $20$, and $59\%$ were female. Students' ethnicity was quite heterogeneous, $69\%$ were Asian, $2\%$ Black, $14\%$ Latinx and $15\%$ White. On average, each work was graded by a random set of five other students. \citealt{Zong2021} produced the peer grading score as the average across different rubrics. As a result, gradings are on a 1-7 continuous scale. The minimum and the maximum observed values were, respectively, 1 and 7. The mean and the median grades were 5.414 and 5.500, respectively, which suggest that data are slightly negatively skewed. To implement the main model, only students who completed at least three assessments were included in the analysis, which resulted in a sample size of 212 students.

\paragraph{Model comparison.}\label{Comparison 1}
Four different models of increased complexity are fitted and compared. In the first model (M1), we only accounted for one student-specific latent variable: the ability and the assessment difficulty level. This model did not consider the effects of graders, such as their systematic biases and reliability levels. Additionally,  M1 assumed that the student's ability was equal across all assessments. This is the more constrained model.
In the second model (M2), we relax our assumptions and consider the graders' effects, such as their systematic bias and reliability levels. To do this, we use a three-dimensional multivariate normal distribution to jointly model the student-specific latent variables, {including $\theta_i$, $\beta_i$ and $\phi^2_i$, $i=1,\dots, N$.}
It is worth noting that fitting M2 is like fitting the reduced model for a single assessment separately (see Section~\ref{CrossSectional}), except that students are assumed to have the same ability level across assessments, i.e., $\theta_{it}=\theta_i$, $i=1,\dots, N$. In the third model (M3), we relax this assumption and allow for variations in students' abilities across assessments {by introducing a fourth student-specific latent variable,  $\eta^2_i$, $i=1,\dots, N$. Under this model, examinee- and grader-specific latent variables, respectively, $(\theta_i, \eta^2_i)$ and $(\beta_i, \phi^2_i)$ are assumed to be independent. This assumption is relaxed in the fourth model (M4) in which the latent variables $\theta_i,\beta_i, \log(\eta^2_i), \log(\phi^2_i)$, $i=1,\dots, N$, are allowed to be correlated. M4 is the main model introduced in Section~\ref{Main}.} 
We also compare these models with the one proposed by \cite{Piech2013} and detailed in Section~\ref{MP}. Under this multiple-assessments setting, we let the difficulty parameter vary across assessments for comparison purposes. 

Model comparison is based on the predictive performance criteria discussed in Section~\ref{ModelComp}. The models are fitted using the prior specifications and posterior procedure discussed in Section~\ref{Bayes}. Grades are on a continuous 1-7 scale, with the midpoint considered the average assessment difficulty. Therefore, we have set the prior distribution for  $\delta_1,\dots,\delta_4 \stackrel{iid}{\sim} N(4,25)$. 
The students are then given an estimate of the true score for each work and a reliability estimate as a grader. 

 \paragraph{Results from the selected Model.}
 Upon graphical inspection of the MCMCs, no mixing or convergence issues were detected, as indicated by $\hat{R}$ values being less than 1.01. The Number of Effective Sample Size was above the cut-off $\hat{N}_{eff}>0.10$ for all the structural parameters \citep{Gelman2013}. The average computational time per chain varies from $64.445$ to $1,594.02$ seconds, respectively, recorded for models M1 and M4. 
 Further details on model diagnostics (e.g., trace plot, $\hat{R}$, $\hat{N}_{eff}$, convergence diagnostic plots), as well as computational time, can be found in Supplementary Materials\footnote{Available through the link \url{https://osf.io/v3ucw/?view_only=aad3bc91cbda43cc9e6c490409323839}}.

Table~\ref{table:1} gives the value of the leave-one-out expected log point-wise density elpd$_{loo}$ and the relative standard error for each model fitted, including the pairwise difference in terms of elpd$_{loo}$ between M4 and each of the other models; in the last column we also report the WAIC \citep{Gelman2013}. The procedure for model comparison showed that M4 provides the best predictive performance. {The slightly better
performance of M4 over M4 in terms of these criteria supports our assumption of the examinee- and grader-specific latent variables being correlated.}

Table~\ref{table:2} shows that the difficulty levels of the assessments are increasing throughout the course. The $95\%$ quantile-based credible intervals of the assessment difficulty parameters are moderately narrow, indicating a low level of uncertainty for these parameters.

The posterior means for the latent variable variances are $\hat{\mu}_3=-1.27$ with a $95\%$ credible interval of $(-1.46,-1.07)$ and  $\hat{\mu}_4=-0.46$ with a 95\% credible interval of  $(-0.51,-0.41)$. This implies that, on average, the variance of the student's ability is smaller than the error variance of the grades they give. In other words, they are more consistent as an examinee than a grader. This seems reasonable, considering that they are not grader experts. Note that these parameters are expressed on a logarithmic scale, meaning that the average variance of the students' proficiency across different assessments is $\exp(\hat{\mu}_3)=0.28$, and, on average, their reliability parameter is $\exp(\hat{\mu}_4)=0.63$. 

Students are moderately homogeneous regarding their mean abilities, as suggested by $\hat{\sigma}_1=0.23$. In contrast, they are more variable in their systematic bias,  $\hat{\sigma}_2=0.35$. In other words, they are, on average, more similar as examinees than as graders. Moreover, students are widely different concerning their consistency across assessments,  $\hat{\sigma}_3=0.66$. Finally, they have slightly less variability concerning the reliability parameters as indicated by $\hat{\sigma}_4=0.32$.

Regarding the dependencies among the latent variables, higher values of students' proficiency are associated with higher consistency values. Indeed, there is evidence of a strong correlation between the first and the second student-specific latent variable, respectively, $\alpha_i$ and $\log(\eta_i^2)$, as suggested by $\hat{\omega}_{13}=-0.86$ and the $95\%$ credible interval of and $(-0.99,-0.76)$. In addition, higher mean bias values predict higher reliability levels. This is evidenced by  $\hat{\omega}_{24}=-0.74$ and the $95\%$ credible interval of $(-0.86,-0.62)$. The estimates of the other correlation parameters do not provide clear evidence about any other dependencies. 
The grader's effect explains, on average, $26.1\%$ of the grading variance, conditioning on the assessment difficulties.

\begin{figure}
    \centering
    \subfigure[]{\includegraphics[scale=0.28]{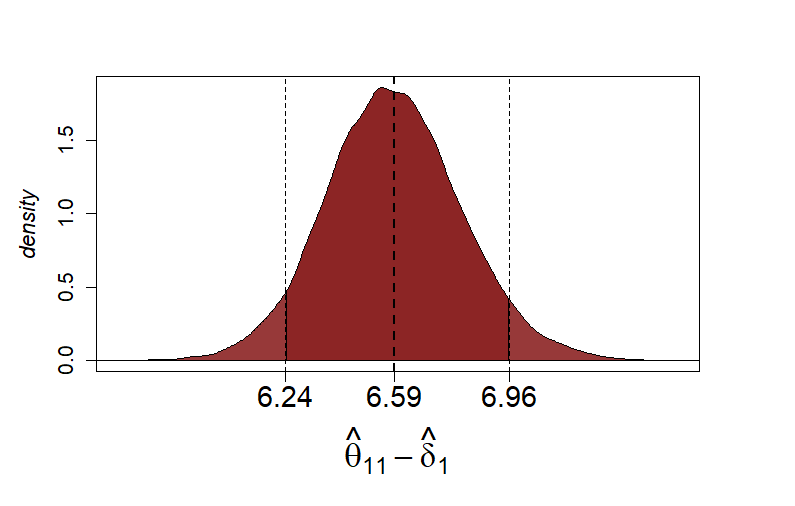}} 
    \subfigure[]{\includegraphics[scale=0.28]{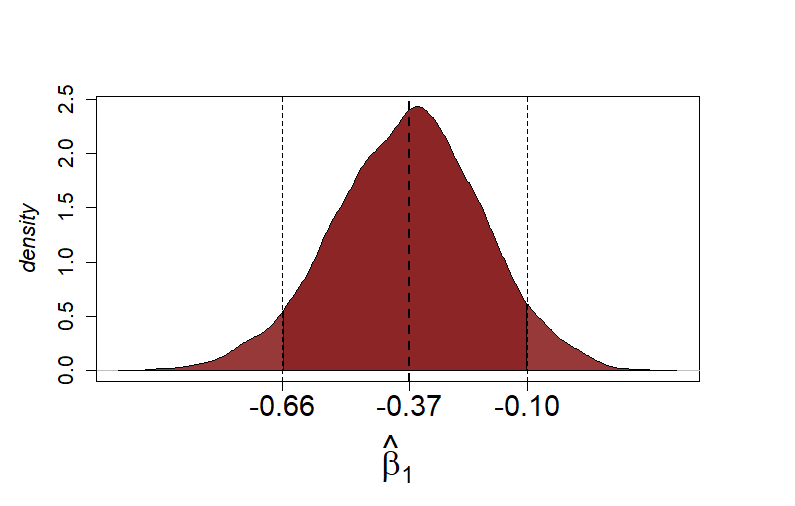}} 
     \subfigure[]{\includegraphics[scale=0.28]{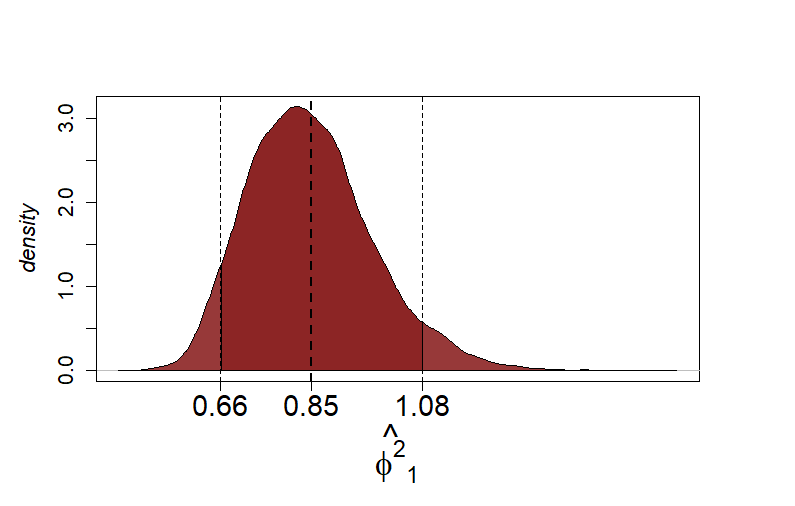}}
    
    \caption{\small Application 1, multiple-assessments example: Posterior distribution of the true score of the first assessment (a), mean bias (b), and reliability (c) of student $i=1$. The black dotted lines indicate the $95\%$ quantile-based credible interval and the posterior mean of each estimated parameter.}
    \label{fig:3}
\end{figure}

At the student-specific level, a score estimate and a $95\%$ quantile-based credible interval may be provided for each assessment to measure the uncertainty. For students' scores, the posterior mean of $\hat{\theta}_{it} - \hat{\delta}_t$ can be used as a point estimate. Additionally, the posterior distributions for both the average bias and the reliability of each grader can be useful in assessing their grading behavior.  If a grader is accurate and reliable, their $\beta_i$ and $\eta_i^2$ values should be close to zero. Conversely, values far from zero indicate biased and unreliable grading behavior. Both parameters are provided with a $95\%$ quantile-based credible interval.
For illustrative purposes, the posterior estimates of the true score $\hat{\theta}_{11} - \hat{\delta}_1$, the mean bias $\beta_1$ and the reliability $\phi_1$ of student $i=1$ are reported in Figure~\ref{fig:3}. On the examinee side, the posterior estimates of the true score suggest that for the first assessment $t=1$, the proficiency level of this student is slightly larger than the average. On the grader side, based on the posterior estimates of $\beta_1$ and $\phi_1^2$, this student is more severe and moderately less reliable than the average (note that $\mu_2=0$ and the posterior mean of $\mu_4$ is $-0.46$ on the log scale).

\begin{table}
\centering
\begin{tabular}{l c c c c c}
\toprule
         Model  & $elpd_{loo}$   & $SE$ & $\Delta elpd_{loo}$ & $SE \Delta$ &  $WAIC$ \\  \midrule
            M4  &  -3751.8       & 49.9 &  --                 &  --     &    7358.18 \\ 
            M3  &  -3770.0       & 49.3 &   $-18.2$           &  $8.5$    &  7382.30 \\
            MP  &  -3937.7       & 52.0 &  $-185.9$           & $24.7$   &   7819.19 \\
            M2  &  -3939.4       & 54.1 &  $-187.6$           & $24.8 $  &   7820.01 \\ 
            M1  &  -4470.7       & 54.2 &  $-718.9$           & $47.1$   &   8939.57 \\ 
         \bottomrule
\end{tabular}
\caption{\small Multiple-assessments example: Four model specifications are compared using a leave-one-out cross-validation approach. The expected log point-wise density value (elpd$_{loo}$) and its respective standard error ($SE$) are reported. The models are given in descending order based on their elpd$_{loo}$ values.  $\Delta elpd_{loo}$ gives the pairwise comparisons between each model and the model with the largest elpd$_{loo}$ (M4), and $SE \Delta$ is the standard error of the difference. The Watanabe–Akaike information criterion (WAIC) is given in the last column for each model.}
\label{table:1} 
\end{table}

\begin{table}
\centering
\begin{tabular}{l c r r}
\toprule
                 & Parameter     &  Post. Mean    &    $95\%$ CI     \\ \midrule
 Assessments     & $\delta_1$    &   $-6.31$       & $(-6.40,-6.22)$    \\
                 & $\delta_2$    &   $-5.39$        & $(-5.47, -5.30)$    \\  
                 & $\delta_3$   &    $-5.36$         & $(-5.44,-5.28)$    \\
                 & $\delta_4$   &    $-4.96$       & $(-5.05,-4.88)$     \\ \midrule
Students         & $\mu_3$    &     $-1.27$          & $(-1.46,-1.07)$    \\
                 & $\mu_4$    &     $-0.46$           & $(-0.51,-0.41)$    \\  
                
                 & $\sigma_1$   &    $0.23$            & $(0.18,0.29)$    \\
                 & $\sigma_2$   &    $0.35$            & $(0.31,0.40)$    \\
                 & $\sigma_3$   &    $0.66$            & $(0.49,0.84)$    \\
                 & $\sigma_4$   &    $0.32$            & $(0.28,0.37)$    \\

                  & $\omega_{12}$   &   $-0.10$            & $(-0.31,0.11)$    \\ 
                 & $\omega_{13}$   &    $-0.86$            & $(-0.99,-0.76)$    \\
                 & $\omega_{14}$   &    $0.17$            & $(-0.05,0.40)$    \\
                 
                 & $\omega_{23}$   &    $-0.07$            & $(-0.32,0.17)$    \\
                 & $\omega_{24}$   &    $-0.74$            & $(-0.86,-0.62)$    \\
                 & $\omega_{34}$   &    $0.12$            & $(-0.15,0.38)$    \\
         \bottomrule
\end{tabular}
\caption[Application 2 - Parameters 95\% credible intervals]{\small Multiple-assessments example: Model M4 estimated structural parameters. The posterior mean (Post. Mean) and the $95\%$ quantile-based credible interval (CI) are reported for each parameter. The parameter $\delta_t$ represents the difficulty level of the assessment $t$; 
$\mu_3$ and $\mu_4$ are the location parameters of the third and the fourth latent variable, respectively; $\sigma_1,\dots,\sigma_4$ are the standard deviations of the latent variables; $\omega_{mn}$ is the correlation parameter between the latent variables $m$ and $n$.}
\label{table:2} 
\end{table}

\subsection{Single Assessment Setting}\label{subsec:real_cross}
The data used for the cross-sectional analysis was obtained from an applied economics undergraduate course at the University of Oviedo, as reported by \citep{Luaces2018}. The sample consisted of 108 students who participated in a double-blind individual peer assessment on an online platform provided by the university. Each coursework was an open-response assessment rated by ten students according to different rubrics on Likert scales of various lengths. For the present analysis, we consider the sum of the ratings given on these other aspects as the final grade. The observed grades ranged from 0 to 12, with a mean of 7.526 and a median of 8.000. 
Further information on the grading procedure might be found in \citealt{Luaces2018}.

\paragraph{Model comparison.}
Four models are fitted and compared. Three are nested models, and the fourth is the model provided by \citet{Piech2013} and discussed in Section~\ref{MP}. 
In the first model (M1), we specify one single student-specific latent variable: the student ability and the assessment difficulty parameter. This model did not consider the effects of graders, such as their systematic biases and reliability levels. In other words, graders' mean bias is fixed to zero, and they are assumed to be equally reliable. This model is the same as the (M1) model detailed in Section~\ref{Comparison 1}, but only with one assessment. In the second model (M2), we let the graders' mean biases be freely estimated. Moreover, we let this second student-specific latent variable correlate with the first one. Indeed, they are assumed to be i.i.d. across students, following a bivariate normal distribution. {In the third model (M3), we relax the assumption of equal reliability across different graders. However, the latent ability $\theta_i$ is assumed to be independent of the other features of the student as a grader (i.e., $\beta_i$ and $\phi^2_i$), for $i=1,\dots, N$. This independence assumption is relaxed in the fourth model (M4) and we allow them to be correlated. M4 is the model presented in Section~\ref{CrossSectional}. } 
The models are fitted using the prior specifications and the posterior procedure discussed in Section~\ref{Bayes}. As with the multiple-assessments example, the prior distribution for the difficulty parameters is set to $N(5.5,25)$. 
The students are then given an estimate of the true score for each assessment and a reliability estimate as a grader. 

 \paragraph{Results for the Selected Model.}
As with the multiple-assessments example, no mixing or convergence issues were detected, as indicated by $\hat{R}$ values less than 1.01.
 The average computational time per chain ranges from $2.7$ to $44.772$ seconds, respectively, recorded from Models M1 and M4. Further details on Model diagnostics (e.g., trace plot, $\hat{R}$), as well as computational time, can be found in the Appendix and an anonymous link\footnote{Available through the link: \url{https://osf.io/v3ucw/?view_only=aad3bc91cbda43cc9e6c490409323839}.}.

 Table~\ref{table:3} indicates that model M4 has the best predictive performance, though its advantage over M3 is very small.  
$\hat{\mu}_3=0.10$ gives the mean graders' reliability level (i.e., the posterior mean of $\eta_i$), and there is considerable variability among them as indicated by $\sigma_3$. 
Indeed, the estimates of $\sigma_3$ on a log scale imply a posterior standard deviation of $\eta_i$ larger than one on the original rating scale.  

Students are very similar in their latent ability, as suggested by the small values of the posterior standard deviation of their abilities, i.e., $\sigma_1$ (see Table~\ref{table:4}). The same extent of variability is estimated concerning their mean biases $\sigma_2$. This implies that students are pretty homogeneous regarding proficiency in doing the assessment and severity in grading their peers. 
The $95\%$ CI for the correlation parameters $\omega_1$ and $\omega_2$ do not suggest a clear relation between the respective latent variables. A positive correlation between graders' bias and their reliability is highlighted by $\omega_{23}$. Nonetheless, the relative credible interval is quite large, implying uncertainty about the correlation size. 
Grader's effects explain, on average, the $16.3\%$ of the grading variance.

\begin{figure}
    \centering
    \subfigure[]{\includegraphics[scale=0.28]{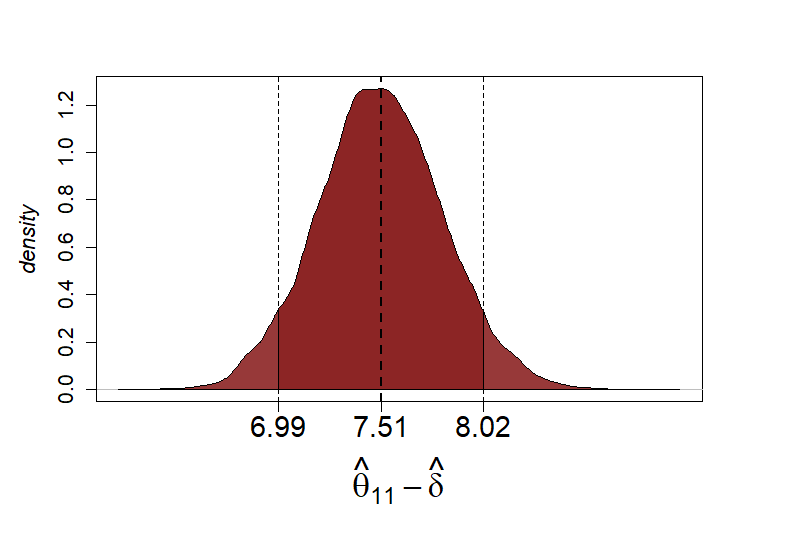}} 
    \subfigure[]{\includegraphics[scale=0.28]{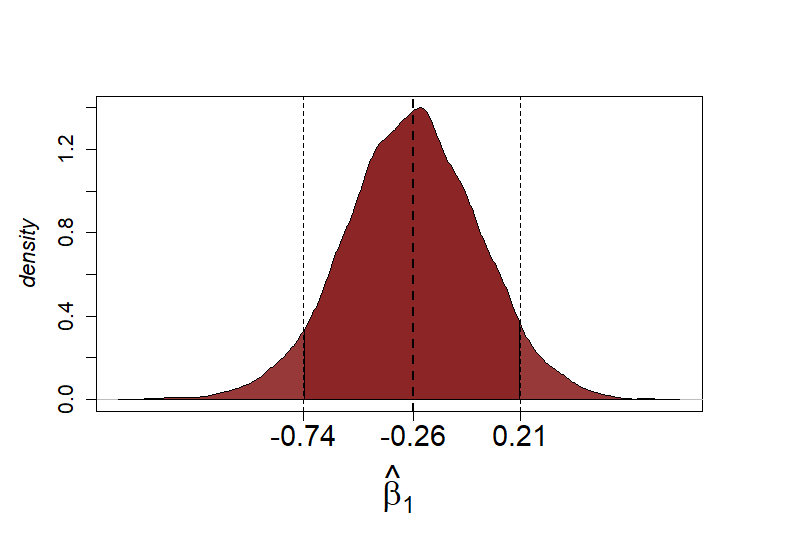}} 
     \subfigure[]{\includegraphics[scale=0.28]{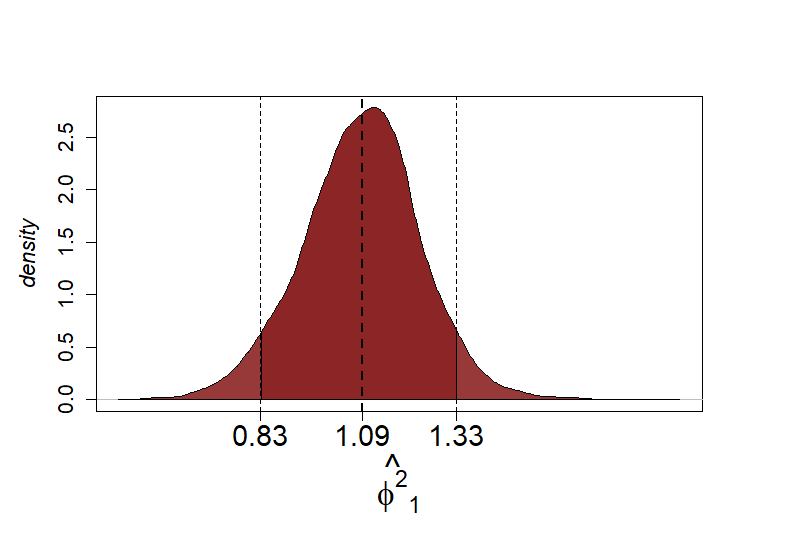}}
    
    \caption{\small Single assessment example: Posterior distribution of the true score (a), mean bias (b),  reliability (c) of student $i=1$. The black dotted lines indicate the $95\%$ quantile-based credible interval and the posterior mean of each estimated parameter.}
    \label{fig:4}
\end{figure}

Each student might receive a true score estimate at the individual level. The posterior mean of $\hat{\theta}_{i} - \hat{\delta}$ might be used as a point estimate for students' true scores. Moreover, the posterior distributions of both the mean bias and the reliability of each grader might be helpful information to assess their grading behavior. 
As an illustration, the posterior estimates of the true score $\hat{\theta}_{11} - \hat{\delta}$, the mean bias $\beta_1$ and the reliability $\phi_1$ of student $i=1$ are reported in Figure~\ref{fig:4}. On the examinee side, the true score's posterior estimates suggest that this student's proficiency level is slightly larger than the average. On the grader side, the posterior estimates of $\beta_1$ and $\phi_1^2$ suggest that this student is moderately more severe than the average in terms of mean bias but average in terms of reliability level (note that $\mu_2=0$ and the posterior mean of $\mu_3$ is $0.10$ on the log scale).

\begin{table}
\centering
\begin{tabular}{l c c c c c}
\toprule
         Model  & $elpd_{loo}$   & $SE$ & $\Delta elpd_{loo}$ & $SE \Delta$ & $WAIC$\\  \midrule
            M4  &   $-1712.6$    & $25.8$    &     0.0         &   0.0      &   3410.29 \\
            M3  &   $-1712.9$    & $25.5$    &   -0.3         &   0.7       &  3410.76 \\ 
            M2  &  $-2271.2$     & $22.4$  &  -558.7         &  19.9       &  4535.97    \\
            M1  &  $-2271.2$     & $22.3$  &  -558.7         &  19.8        & 4536.05  \\ 
            MP  &  $-2283.9 $     & $22.8$  &  -571.4         &  20.8        & 4557.39 \\     
         \bottomrule

\end{tabular}
\caption{\small Single assessment example: Four model specifications are compared using a leave-one-out cross-validation approach. The expected log point-wise density value (elpd$_{loo}$) and its respective standard error ($SE$) are reported. The models are given in descending order based on their elpd$_{loo}$ values.  $\Delta elpd_{loo}$ gives the pairwise comparisons between each model and the model with the largest elpd$_{loo}$ (M4), and $SE \Delta$ is the standard error of the difference. The Watanabe–Akaike information criterion (WAIC) is given in the last column for each model.}
\label{table:3} 
\end{table}

\begin{table}
\centering
\begin{tabular}{l c r r}
\toprule
                 & Parameter     &  Post. Mean    &    $95\%$ CI     \\ \midrule
 Assessment     & $\delta$    &   $-7.19$       & $(-7.33, -7.07 )$   \\ \midrule
Students         & $\mu_3$    &     $0.10$          & $(0.05,0.15)$    \\
                 
                 & $\sigma_1$   &    $0.54$          & $(0.43,0.65)$      \\
                 & $\sigma_2$   &    $0.53$            & $(0.42,0.65)$    \\
                 & $\sigma_3$   &    $0.15$            & $(0.03,0.24)$    \\

                  & $\omega_{12}$   &   $-0.06$            & $(-0.31,0.18)$   \\ 
                 & $\omega_{13}$   &    $0.29$            & $(-0.08,0.66)$   \\
                 & $\omega_{23}$   &    $-0.52$            & $(-0.87,-0.15)$    \\
         \bottomrule
\end{tabular}
\caption{\small Application 2, single assessment example: Model M4 estimated structural parameters. The posterior mean (Post. Mean) and the $95\%$ quantile-based credible interval (CI) are reported for each parameter. The parameter $\delta$ represents the difficulty level of the assessment; $\mu_3$ is the location parameter of the third latent variable; $\sigma_1,\dots,\sigma_3$ are the standard deviations of the latent variables; $\omega_{mn}$ is the correlation parameter between the latent variables $m$ and $n$. }
\label{table:4} 
\end{table}

\section{Discussions}\label{sec:discussion}

This paper presents a new modeling framework for peer grading data, which introduces latent variables to capture the dependencies in the data from the network structure of peer grades and the dual role of each student as an examinee and a grader. The statistical inference uses a Bayesian method, and an algorithm based on the No-U-Turn Hamiltonian Monte Carlo sampler was developed. The proposed model was applied to two real-world peer grading datasets, one with a single assessment and the other with four. The results showed that the proposed model had superior prediction performance in real-world applications and that the MCMC did not suffer from mixing or convergence issues.

The current work also has some limitations. First, the peer grades in the applications in Section~\ref{sec:real} 
are bounded, which may cause ceiling and floor effects, as the variability of student performance is no longer measurable when they receive a very high or low score. However, the proposed model is based on several normal assumptions, which fail to capture such phenomena. To model bounded grades, we may add a nonlinear transformation to the right-hand side of \eqref{eq:1} to ensure $Y_{ig}$ to be bounded. 
 
Second, it is not easy to verify the assumptions about the latent variables in our model and further validate their interpretation, as we cannot observe the latent variables. Without additional information, it is hard to disentangle different assumptions about the latent variables and verify them separately. We can only check whether the model-implied distribution for the observed data fits its empirical distribution (e.g., using Bayesian LOO and WAIC) and use it to compare different models.  Using this approach, we can only tell that the assumptions of one model as a whole are 
more sensible than those of the other. 
To further verify the assumptions of our model, we may collect data with both peer and instructor grades. The instructor grades may be used as the underlying truth to check some specific assumptions in our model. 

The current work can be further extended in several directions. First, in formative assessment settings, people are often interested in the growth of students over multiple assessments during a course. Therefore, it may be useful to extend the proposed model to a longitudinal setting and develop a latent growth curve model for peer grading. To explore this direction, we have considered a simple extension of the proposed model and performed a small simulation study in Appendix~\ref{Extensions}. This model assumes the true score $\theta_{it}$ to follow a latent growth curve model. While this model performed well in the simulation, it may be over-simplified for real-world settings. In practice, student characteristics as a rater and the difficulty levels of the assessments may also change over time. Simultaneously modeling all these changes may result in model identification issues. We leave this problem for future investigation.


Second,  additional context information, such as student- and classroom-related factors, is often available in formative assessment settings. Such information is useful in explaining and predicting
each student's performance both as an examinee and as a rater. In this regard, we believe it is useful to extend the framework of explanatory item response models \citep{Kim_2020,wilson2004} to the current setting to include context information as covariates.

Third, the reliability of the peer grading system based on the proposed model is worth investigating. This may be done by adapting the generalizability theory \citep{Brennan2001,Brennan2010}, originally established under the traditional cross-classified random effects models, to the current model. With the new generalizability theory, we may evaluate the reliability of the system from different perspectives (e.g., examinees, raters, and assessments). 

Fourth, many real-world peer grading systems involve ordinal peer grades. The proposed model may be extended to ordinal data by replacing the linear model \eqref{eq:1} with a generalized linear model. One possible formulation is given in Appendix~\ref{app:ord}, which still mimics the proposed model but replaces 
\eqref{eq:1} with a partial credit model form \citep{masters1982rasch}. Alternative models also may be available, such as one based on the graded response model \citep{Samejima1969estimation}. The suitability of these models for peer grading remains to be studied through a theoretical investigation and numerical studies based on 
simulated and real data. We leave it for future investigation.


Finally, it should be noted that although the Bayesian approach allows for statistical inferences, it can be time-consuming to compute. The high computational cost is due to the proposed model's high dimensionality, which depends on the number of students and assessments. To make the proposed method scalable for large-scale applications, like MOOC data, advanced computational methods for Bayesian inference, such as stochastic gradient MCMC algorithms, should be explored \citep{nemeth2021stochastic}.

\section*{\large  Acknowledgments }
We are sincerely grateful to Professors Oscar Luaces and Christian Schunn 
for sharing real datasets in Section~\ref{sec:real}.

 \bigskip
 \bigskip
 
\appendix
\noindent
{\Large{\textbf{Appendix}}}  

\counterwithin{figure}{section}
\counterwithin{table}{section}
\numberwithin{equation}{section}

\section{Computational details}
To resolve the convergence issue and make the MCMC mix well, we used a non-centred reparametrization for the multivariate normal distribution \citep{Gelman2013,Betancourt2013,Papaspiliopoulos2007}. We express the distribution of the vector of student-specific latent variables through an affine transformation, such that: 
\begin{eqnarray}\label{eq:4}
    \left(\alpha_{i}, \beta_{i}, \log(\eta_{i}^2), \log(\phi_{i}^2)\right) &=& \bm{\mu} + \textbf{S}\left(\textbf{L}\bm{\gamma}_i \right). \nonumber
\end{eqnarray}
Here $\textbf{L}$ is the Cholesky factor of the correlation matrix $\bm{\Omega}=\textbf{L}\textbf{L}'$ and $\bm{S}$ is the diagonal matrix of the standard deviation of the latent variables, $\bm{S}= diag(\sigma_j)$, $j=1,\dots,4$; note that $\bm{\Sigma} = \bm{S} \bm{\Omega} \bm{S}$. The element of the four-dimensional vector $\bm{\gamma}_i \in \mathbb{R}^4$ are i.i.d. following a standard normal distribution, $\gamma_{i,1},\dots,\gamma_{i,4}\stackrel{iid}{\sim}N(0,1)$, $i=1,\dots,N$. Notice that, as stated above, $\mu_1=\mu_2=0$ for identifiability purposes. \\
Under this inference procedure, each parameter and student-specific latent variables might be provided with a posterior point estimate, e.g., the posterior mean, and an interval estimate, e.g., a $95\%$ quantile-based credible interval. The latter might be seen as an uncertainty measure of the estimated parameter; broader intervals suggest more uncertainty about the values of the parameters, whereas narrower intervals reflect less uncertainty about their values. 

\section{Extension to Latent Growth Curve Model}\label{Extensions}

\subsection{Model Specification}
When sufficient assessments are given over time, evaluating students' growth during that period may be interesting. This can be done using Latent Growth Curve (LGC) modeling  \citep{bollen2006latent}. For example, a linear latent curve unconditional model can expand the  structural model \eqref{eq:truescore} for true score $\theta_{it}$ by assuming
\begin{equation}
 \theta_{it} \sim N(\alpha_{i0} + \lambda_t \alpha_{i1}, \eta_i^2),
\end{equation} 
where $\alpha_{i0}$, $\alpha_{i1}$ and $\eta_i^2$ are student-specific latent variables, and $\lambda_t$, $t=1, ..., T$ are a pre-specified coding of time. 
The linear function $\alpha_{i0} + \lambda_t \alpha_{i1}$ can be interpreted as the latent trajectory of student $i$. The coding of time can be chosen based on the time when the assessments are given. In the special case when the assessments are given at equally spaced intervals, 
we can set $\lambda_t = t-1, t=1, 2, ..., T$. 
We keep the model for $\tau_{igt}$ unchanged. The student-specific latent variables now include  $(\alpha_{i0}, \alpha_{i1}, \beta_{i}, \eta_{i}^2,  \phi_{i}^2)$. In line with our assumptions in the main model, we  assume that 
$(\alpha_{i0}, \alpha_{i1}, \beta_{i}, \log(\eta_{i}^2),  \log(\phi_{i}^2))$, $i=1, ..., N$ are i.i.d., and follow a multivariate normal distribution. 

This model can be further extended to capture nonlinear trajectories. For example, a quadratic latent curve model may be assumed for $\theta_{it}$ by assuming 
\begin{equation}
 \theta_{it} \sim N(\alpha_{i0} + \lambda_t \alpha_{i1} +\lambda_t^2 \alpha_{i2}, \eta_i^2),
\end{equation}
where $\alpha_{i0}$, $\alpha_{i1}$, $\alpha_{i2}$ and $\eta_i^2$ are student-specific latent variables, and $\lambda_t$, $t=1, ..., T$ are still a pre-specified coding of time. The random second-order coefficient $\alpha_{i2}$ is the rate of
change in the linear component along
time and represents the acceleration in growth for student $i$' ability \citep{bollen2006latent,Biesanz2004}. The joint model for student-specific latent variables can be extended accordingly. 

The prior specifications discussed in Section~\ref{sec:model} for the unknown parameters, $\delta_1,\dots,\delta_T$ and $\bm{\mu},\bm{\Sigma}$, might be consistently adopted for the current model. The same procedures and reparametrization used for the posterior computation introduced above might be freely applied here for the multivariate normal distribution. The Bayesian model comparison procedure discussed in Section~\ref{sec:model} might be used to compare the models under both the main and LGC framework. As for the previous models, each parameter and student-specific latent variables might be provided with both a posterior point estimate, e.g. the posterior mean, and an interval estimate, e.g. a $95\%$ quantile-based credible interval. Indeed it might be seen as an uncertainty measure of the relative estimated parameter.

\subsection{A Simulation under the LGC Model}
{Homoscedastic LGC models typically require at least three time-points per individual \citep[e.g.,][]{Curran_2010}, whereas, our proposal concerns a heteroscedastic setting that allows some variance terms to be individual-specific and implies a larger number of parameters. Given the relatively small number of assignments per student (at most four) and the rather small sample size $N=212$, the LGC model might not be suitable for the real data analyzed in Section \ref{subsec:real_Long}. }
We provide a simulated example based on the above LGC model an illustrative example. 

\paragraph{Data generation.} 
We generated a dataset from the linear LGC Model in which a sample of $N=100$ students are assigned $T=6$ assessments. For each assessment, the work of each student is graded by a random subset of other $|S_{it}|=3$ students.  
The following values are fixed for the structural parameters of the model:
$\mu_\delta=0$, $\sigma_\delta=1$, $\bm{\mu}=0$, $\bm{\Omega}=\bm{I}$ is a $5$-dimensional identity matrix, $\bm{S}=diag(1,0.1,1,0.2,0.2)$, see Table~\ref{table:5}. 

\paragraph{Estimated parameters.}
All the considerations on the prior and computational aspects discussed in the main text are consistently followed here. 
The graphical inspection of the MCMCs does not suggest any mixing or convergence issues which is consistent with the low values of $\hat{R}<1.01$. 
 The average computational time per chain recorded is $5,293.08$ seconds\footnote{More details on computational time and model diagnostic are available through the link: \url{https://osf.io/v3ucw/?view_only=aad3bc91cbda43cc9e6c490409323839}}.
 
 The estimates of the structural parameters are reported in Table~\ref{table:5}. All the true values are included in the $95\%$ credible intervals, even though these posterior intervals are considerably large. This uncertainty might be due to the small sample size $N=100$. The only exception is the correlation parameter  $\omega_{34}$ which $95\%$ credible interval does not include the true value (even if it is the difference between the upper bound of the interval and the true value is practically negligible). More replications might shed light on these aspects. 
 
\begin{figure}
    \centering
    \subfigure{\includegraphics[scale=0.28]{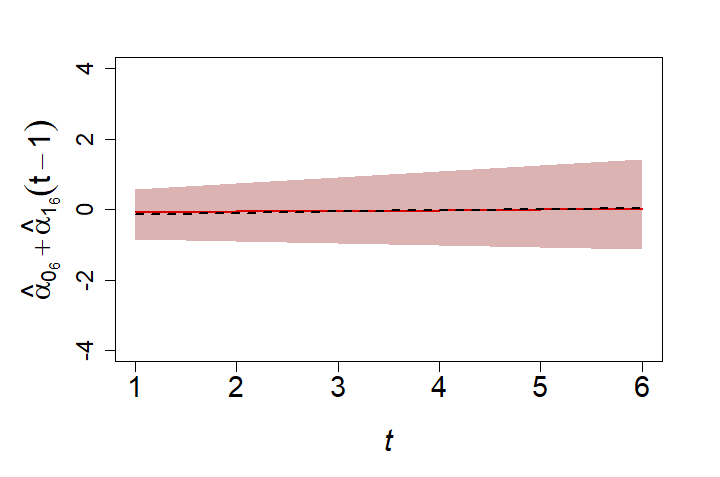}} 
    \subfigure{\includegraphics[scale=0.28]{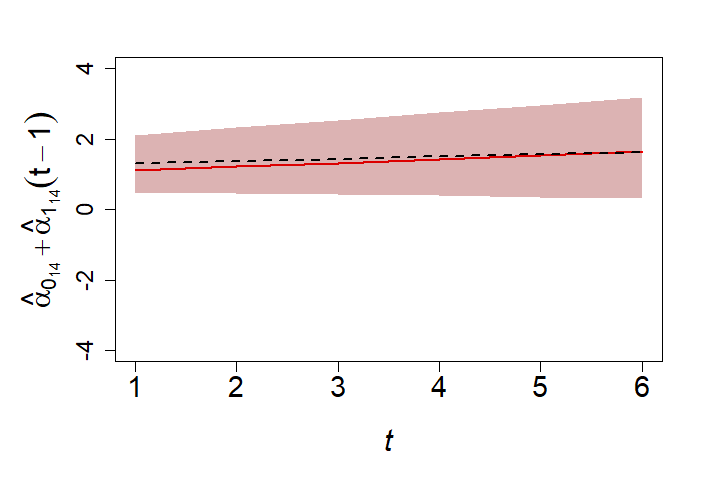}} 
     \subfigure{\includegraphics[scale=0.28]{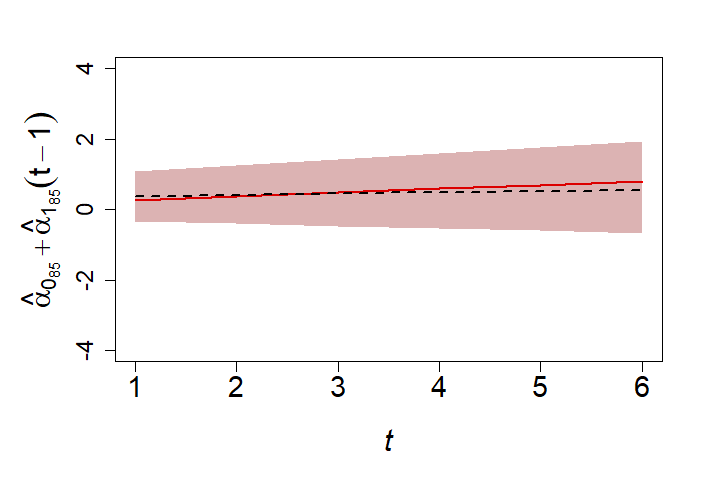}}
    \subfigure{\includegraphics[scale=0.28]{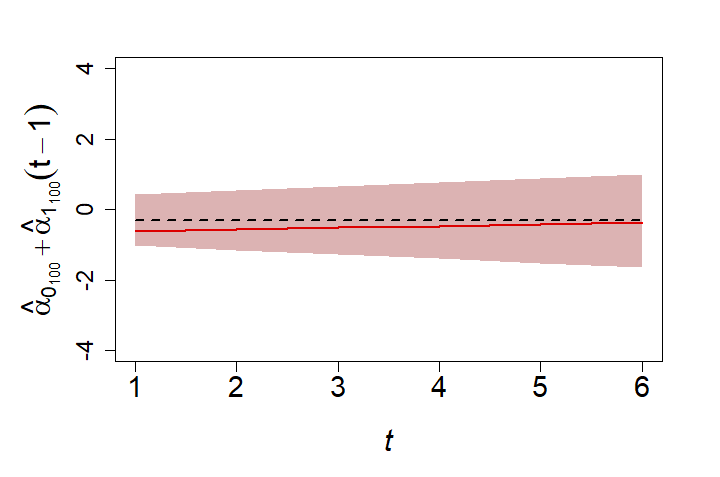}}
    
    \caption{\small Estimated linear latent growth of a random sample of four students. The red and the black dotted lines are the true linear growth and the estimated one, respectively; the red bands are the $95\%$ credible intervals of this trend.}
    \label{fig:5}
\end{figure}

 Under this model, along with the estimates of the true grade and the grader's effects (i.e., the mean bias and the reliability), each student might be provided with the estimate of his/her latent linear growth trend. For illustrative purposes, we plot the estimated growth of four different students in Figure~\ref{fig:5}. Note that the $95\%$ credible intervals of the student-specific trend $\alpha_{0i}+\alpha_{1i}(t-1)$ are proportional to the values of $t$. This is because the $0.05$ and the $0.95$ quantiles of the posterior of $\alpha_{1i}$ are multiplied by this covariate.

\begin{table}
\centering
\begin{tabular}{l c r r r}
\toprule
                 & Parameter     &  True Value    &  Post. Mean    &    $95\%$ CI     \\ \midrule
 Assessments     & $\mu_\delta$    &    $0.00$             &   $0.05$       & $(-3.72, 3.80 )$   \\ 
                 & $\sigma_\delta$   &    $2.00$             &   $3.59$       & $(1.95, 6.35)$   \\

                & $\delta_1$    &    $2.52$             &   $2.48$       & $(2.18, 2.78)$   \\ 
                & $\delta_2$    &    $-1.72$             &   $-1.74$       & $(-3.04, -0.40 )$   \\ 
                & $\delta_3$    &    $-1.01$             &   $-0.95$       & $(-3.53,  1.67  )$   \\ 
                & $\delta_4$    &    $1.39$             &   $1.01$       & $(-2.84,  4.90 )$   \\ 
                & $\delta_5$    &    $-3.45$             &   $-3.55 $       & $(-8.69,  1.65 )$   \\ 
                & $\delta_6$    &    $3.30$             &   $3.27$       & $(-3.15, 9.75)$   \\  \midrule

Students         & $\mu_2$       &       $0.00$         &  $-0.05$          & $(-1.32,1.26)$    \\
                 & $\mu_4$       &       $0.00$       &  $-0.01$          & $(-0.09,  0.06)$    \\
                 & $\mu_5$       &       $0.00$        &  $0.01$          & $(-0.03,0.05)$    \\
                           
                 & $\sigma_1$    &        $1.00$       &    $1.02$          & $(0.85, 1.20)$      \\
                 & $\sigma_2$    &        $0.01$    &    $0.07$            & $(0.01, 0.15)$    \\
                 & $\sigma_3$    &        $1.00$     &    $1.03$            & $(0.91, 1.17)$    \\
                 & $\sigma_4$    &         $0.20$      &    $0.17$            & $( 0.05, 0.28)$    \\
                 & $\sigma_5$    &         $0.20$    &    $0.19$            & $( 0.16, 0.23)$    \\

                  & $\omega_{12}$&    $0.00$              &   $0.11 $            & $( -0.43, 0.66)$   \\ 
                  & $\omega_{13}$ &   $0.00$               &    $0.11$            & $(-0.09, 0.31)$   \\
                  & $\omega_{14}$ &   $0.00$               &    $-0.13$            & $(-0.53, 0.30)$    \\
                  & $\omega_{15}$&    $0.00$              &   $0.10$            & $(-0.13, 0.33)$   \\ 
             
                 & $\omega_{23}$ &    $0.00$              &    $-0.30$            & $(-0.76, 0.21)$   \\
                 & $\omega_{24}$ &    $0.00$              &    $0.13$            & $(-0.49, 0.69)$    \\
                 & $\omega_{25}$ &    $0.00$              &    $0.03 $            & $(-0.50, 0.54)$   \\
                
                 & $\omega_{34}$ &    $0.00$              &    $-0.40$            & $( -0.74,-0.03)$    \\
                 & $\omega_{35}$ &    $0.00$              &    $0.13$            & $(-0.07,0.34)$    \\

                 & $\omega_{45}$ &    $0.00$              &    $0.33 $            & $(-0.10,0.70)$    \\ 
         \bottomrule
\end{tabular}
\caption{\small Estimated structural parameters. The true value, the posterior mean (Post. Mean) and the $95\%$ quantile-based credible interval (CI) are reported for each parameter. The parameter $\delta$ is the difficulty level of the assessment; $\mu_2$,$\mu_4$ and $\mu_5$ are the location parameters of the second, the fourth and the fifth latent variable, respectively; $\sigma_1,\dots,\sigma_5$ are the standard deviations of the latent variables; $\omega_{mn}$ is the correlation parameter between the latent variables $m$ and $n$.}
\label{table:5} 
\end{table}

\section{Prior Sensitivity Analysis}
We performed a prior sensitivity analysis to investigate the impact of the prior on final model estimates \citep{Gelman2013}. As discussed by \citet{Gelman_2006_BA}, inferences might be very sensitive to the choice of the prior distribution for hierarchical variance parameters and useful information might come from tailored stimulative studies. As a preliminary analysis, we fit the main model on different generated data sets comparable to the real data set analysed in Section \ref{subsec:real_Long}. For each data set, we alternatively fit the main model under three different prior specifications resulting in three different scenarios. We compare the respective estimates through the root mean square error (RMSE) and the mean absolute error (MAE). 

\paragraph{Data generation.} 
We generated $R=10$ independent datasets from the Main Model in which a sample of $N=100$ students are assigned $T=4$ assessments. For each assessment, each student's work is graded by a random subset of other $|S_{it}|=3$ students.  
The following values are fixed for the structural parameters of the model:
$\mu_\delta=0$, $\sigma_\delta=1$, $\bm{\mu}=0$, $\bm{\Omega}=\bm{I}$ is a $4$-dimensional identity matrix, $\bm{S}=diag(1,1,0.2,0.2)$. 

\begin{figure}
    \centering
    \includegraphics[scale=0.6]{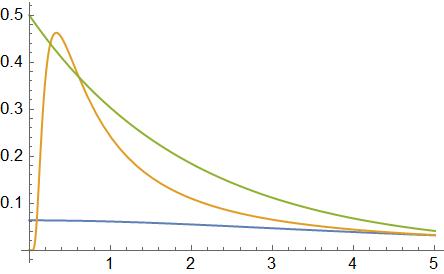}
    
    \caption{\small Priors placed under different scenarios on $\sigma_1, \dots,\sigma_4, \mu_3,\mu_4$. The blue, orange and green solid lines indicate, respectively, the half-Cauchy, the inverse-gamma and the exponential priors.}
    \label{fig_priors}
\end{figure}

\paragraph{Scenarios and Priors}
We place three different priors on the scale parameters $\sigma_1, \dots,\sigma_4$ and on the means $\mu_3,\mu_4$. Under the first scenario, they are assigned a half-Cauchy as recommended by \citet{Gelman2013}, $\sigma_1, \dots,\sigma_4 \stackrel{iid}{\sim} half-Cauchy(0,5)$ and  $\mu_3,\mu_4 \stackrel{iid}{\sim} half-Cauchy(0,5)$. This class of priors are referred to as "weekly informative" by \citet{Gelman_2006_BA} because of the gentle slope of their tails which can let the data dominate the posterior if the likelihood is strong in that
region. In the second scenario, we place an inverse-gamma on those parameters,  $\sigma_1, \dots,\sigma_4 \stackrel{iid}{\sim} InvGamma(0.5,0.5)$ and  $\mu_3,\mu_4 \stackrel{iid}{\sim} InvGamma(0.5,0.5)$. Under the third scenario, they are assigned an exponential prior,  $\sigma_1, \dots,\sigma_4 \stackrel{iid}{\sim} \exp(0.5)$ and  $\mu_3,\mu_4 \stackrel{iid}{\sim} \exp(0.5)$. As shown by Figure \ref{fig_priors}, the probability density is more spread and diffused under the first scenario as a result of the half-Cauchy prior specification. 

\paragraph{Results}
Inferences are the same across different scenarios, suggesting a robustness of the model to different prior specifications for the scale parameters. Table \ref{Tab_RMSEA} gives the RMSE and the MAE for the structural parameters and the true scores (i.e., $\theta_{it}-\delta_t$). The RMSE and the MAE of the estimates of the true scores are, on average, 1.99 and 1.59 under all the scenarios. The RMSE and the MAE of the aggregated score using the mean to derive the final grade for each student's work \citep{Sajjadi2015, Reily2009} are, respectively, 2.46 and 1.98. This suggests that our proposal might consistently mitigate graders' systematic bias and unreliability. \\

\begin{table}
\centering
        \begin{tabular}{l c c c c c c}
           \toprule
           Parameter & \multicolumn{6}{c}{Scenarios} \\
            \midrule 
                      & \multicolumn{2}{c}{\textit{1}} &  \multicolumn{2}{c}{\textit{2}} & \multicolumn{2}{c}{\textit{3}}\\
           \midrule
           &           RMSE & MAE &    RMSE & MAE &  RMSE & MAE \\
            \midrule 
         
         $\delta_1$   &  0.243 & 0.195 & 0.244 & 0.196 &  0.244 & 0.196      \\
         $\delta_2$   &  0.182 & 0.138 & 0.182 & 0.137 &  0.183 & 0.138      \\
         $\delta_3$   &  0.159 & 0.138 & 0.161 & 0.142 &  0.159 & 0.138      \\
         $\delta_4$   &  0.131 & 0.095 & 0.131 & 0.096 &  0.131 & 0.095      \\   \midrule

        $\mu_3$    &     1.018 & 1.015 &  1.012  &  1.009 &  0.994  & 0.991    \\
        
        $\mu_4$      &    1.001     & 0.997   &    0.999      &   0.996    &    1.029    &   1.023      \\

        $\sigma_1$   &    0.150     &  0.126   &   0.127      &  0.108    &   0.109    &   0.090    \\
        $\sigma_2$   &     0.145    &  0.124   &    0.105     &   0.088   &   0.133    &   0.102    \\
        $\sigma_3$   &    0.834     &   0.825  &   0.772      &   0.760   &   0.840    &   0.826    \\
        $\sigma_4$   &    0.855     & 0.841    &    0.811      &  0.806    &   0.849    &  0.847     \\

         $\omega_{12}$ & 0.124 &  0.101     & 0.124 &  0.101    &   0.122 &   0.099    \\
         $\omega_{13}$ & 0.220 &  0.184     & 0.162 &  0.136    &   0.215 &   0.181    \\
         $\omega_{14}$ & 0.246 &  0.224     & 0.235 &  0.210    &   0.247 &   0.226    \\
         $\omega_{23}$ & 0.136 &  0.085     & 0.116 &  0.068    &   0.135 &   0.084    \\
         $\omega_{24}$ & 0.275 &  0.214     & 0.262 &  0.202    &   0.276 &   0.215    \\
         $\omega_{34}$ & 0.119 &  0.078     & 0.078 &  0.054    &   0.116 &   0.082    \\  \midrule

          True score & 1.995   & 1.597   & 1.994 & 1.596 & 1.995 & 1.597 \\  

 \bottomrule
        \end{tabular}
        \label{Tab_RMSEA}
\caption[]{ \small  Root Mean Square Error (RMSE) and Mean Absolute Error (MAE) related to students' true scores and structural parameters under different scenarios across 10 independent data sets. }
    \end{table}

\section{An Extension to Ordinal Peer Grades}\label{app:ord}

The models in this paper are for continuous grades, while grades in many peer grading systems are on an ordinal scale. In what follows, we discuss how the proposed model may be extended to ordinal grades. We consider a formulation based on the partial credit model \citep{masters1982rasch} under the same setting as our main model in Section~\ref{Main} except for the grades being ordinal. More specifically, suppose that  $Y_{igt} \in \{1,\dots,K\}$. 
For each $k = 2, ..., K$, we assume that 
\begin{equation}\label{eq:pcm}
P(Y_{igt} = k | \theta_{it},\beta_g,\phi_g > 0,\boldsymbol{\delta}_t, Y_{igt} \in \{k-1, k\}) = \frac{\exp{((\theta_{it} + \beta_g - \delta_{t,k-1})/\phi_{g})}}{1+\exp{((\theta_{it} + \beta_g - \delta_{t,k-1})/\phi_g)}},
\end{equation}
where  $\boldsymbol{\delta}_t=(\delta_{t1},\dots, \delta_{t, K-1})^\top$ contains the item-specific parameters and the rest of the variables can be interpreted similarly as in Section~\ref{Main}. More specifically,  $\theta_{it}$ may be interpreted as student $i$'s true score for assessment $t$.  The larger the value of  $\theta_{it}$, the more likely that $Y_{igt}$ takes a value in a higher category. The variable $\beta_g$ can still be interpreted as rater $g$'s bias, as raters with a larger  $\beta_g$ value tend to give a higher grade on average. In addition, 
$\phi_g$ still indicates rater $g$'s reliability. When $\phi_g$ goes to infinity and the rest of the parameters remain fixed, the probability in \eqref{eq:pcm} will converge to 0.5, and thus, the probability of $Y_{igt} = k$ will converge to $1/K$, for each category $k$, regardless what the true score $\theta_{it}$ is. In other words, the grade $Y_{igt}$ becomes a purely random guess. On the other hand, when $\phi_g$ goes to zero, the distribution of $Y_{igt}$ will concentrate on one of the categories. 

Similar to the model in Section~\ref{Main}, we can assume $\theta_{it}$ to follow \eqref{eq:truescore}, and further set priors for the student-specific latent variables, as well as the rest of the model parameters. Bayesian inference can then be performed.  

\bigskip

\bibliographystyle{apacite}

\bibliography{reference}

\end{document}